# Magnetoresistance anomaly during the electrical triggering of a metal-insulator transition

Pavel Salev[1], Lorenzo Fratino[2,3], Dayne Sasaki[4], Soumen Bag[2], Yayoi Takamura[4],
Marcelo Rozenberg[2], Ivan K. Schuller[5]

*[1]Department of Physics & Astronomy, University of Denver, Denver, CO*
*[2]Université Paris-Saclay, CNRS Laboratoire de Physique des Solides, 91405, Orsay, France*
*[3]Laboratoire de Physique Théorique et Modélisation, CNRS UMR 8089, CY Cergy Paris Université, 95302 Cergy-Pontoise Cedex, France*
*[4]Department of Materials Science and Engineering, University of California Davis, Davis, CA*
*[5]Department of Physics and Center for Advanced Nanoscience, University of California San Diego, La Jolla, CA*

Phase separation naturally occurs in a variety of magnetic materials and it often has a major impact on both electric and magnetotransport properties. In resistive switching systems, phase separation can be created on demand by inducing local switching, which provides an opportunity to tune the electronic and magnetic state of the device by applying voltage. Here we explore the magnetotransport properties in the ferromagnetic oxide $(La,Sr)MnO_3$ (LSMO) during the electrical triggering of an intrinsic metal-insulator transition (MIT) that produces volatile resistive switching. This switching occurs in a characteristic spatial pattern, i.e., the formation of an insulating barrier perpendicular to the current flow, enabling an electrically actuated ferromagnetic-paramagnetic-ferromagnetic phase separation. At the threshold voltage of the MIT triggering, both anisotropic and colossal magnetoresistances exhibit anomalies including a large increase in magnitude and a sign flip. Computational analysis revealed that these anomalies originate from the coupling between the switching-induced phase separation state and the intrinsic magnetoresistance of LSMO. This work demonstrates that driving the MIT material into an out-of-equilibrium resistive switching state provides the means to electrically control of the magnetotransport phenomena.

## I. INTRODUCTION

Resistive switching electronics and spintronics emerged as two leading frameworks for the development of scalable and energy-efficient memories and signal processing devices for next generation information technologies [1–3]. In resistive switching systems, an electric stimulus, voltage or current, programs the material's resistivity either in a volatile or nonvolatile way [4]. In spintronics, electrical signals are used to manipulate and probe the magnetic degrees of freedom [5]. Combining resistive switching and spintronic functionalities in a single device presents an exciting opportunity to bring together the advantages of charge- and spin-based electronics, enriching the design space for practical applications, and to further the basic understanding of the interactions between electrical and magnetic phenomena in materials.

A common mechanism of the nonvolatile resistive switching is the electrically induced ionic migration [6]. The application of voltage to a metal-insulator-metal heterostructure can drive ion diffusion from the metal electrodes into the insulator layer [5] or cause the ion rearrangement within the insulator [7] leading to a change of the heterostructure's resistance. When a magnetic material is incorporated into an ionic migration device, inducing the resistive switching can enable the electrical control of magnetic properties such as anisotropic and tunneling magnetoresistances [8–14], saturation magnetization [15–17], exchange bias [18,19], and magnetic domain configuration [20,21].

The application of an electric stimulus to a material featuring a metal-insulator transition (MIT) can produce a large resistance change without the ionic migration [4]. Typically, the applied voltage/current drives the MIT material across the phase transition either by Joule heating or by field-induced carrier doping [22–24]. Electrical triggering of the MIT results in a volatile resistive switching, i.e., a switching that automatically resets upon the removal of the electrical stimulus. Volatile MIT switching is actively being pursued in applications including selectors in crossbar arrays [25,26], rf and optoelectronics switches [27–29], and spiking oscillators for neuromorphic [30–33] and stochastic [34,35] computing. Because of the volatility, the MIT switching is not directly applicable in memories, although several works have demonstrated nonvolatile memristive functionalities in the MIT devices by utilizing the hysteretic first order transitions [36–38], ramp-reversal memory [39,40], and artificial heterostructures [41,42].

While the electrical properties in the MIT switching devices have been studied extensively, the magnetic functionalities remain largely unexplored despite the fact that many MIT materials also have magnetic ordering [43]. During the electrical MIT triggering, a different electronic and magnetic phase is locally injected into the otherwise homogeneous material, often due to the formation of a percolating filament [44] or transverse barrier [45]. It can be expected that the electrically induced phase separation can influence magnetic ordering similar to the naturally occurring phase separation in magnetic materials [46]. Electrical control of the phase separation by the MIT switching can enable the electrical manipulation of magnetic properties without altering the material's chemical composition as in the case of ionic-migration systems. Recently we reported magneto-optical studies of the uniaxial magnetic anisotropy emergence in a ferromagnetic MIT oxide $(La,Sr)MnO_3$ (LSMO) when the applied voltage induces volatile resistive switching [47]. In this work, we focus on the magnetotransport properties during the electrical triggering of MIT in LSMO. We observed strong anomalies both in anisotropic and colossal magnetoresistances (AMR and CMR), including a large change in magnitude and sign flip of the magnetoresistance. We attributed these magnetotransport anomalies to the coupling between the switching driven metal/insulator/metal phase separation and intrinsic magnetoresistance of LSMO. Our findings provide a unique approach to control the spin-dependent transport in the MIT materials that are driven out of equilibrium by strong electrical stimuli.

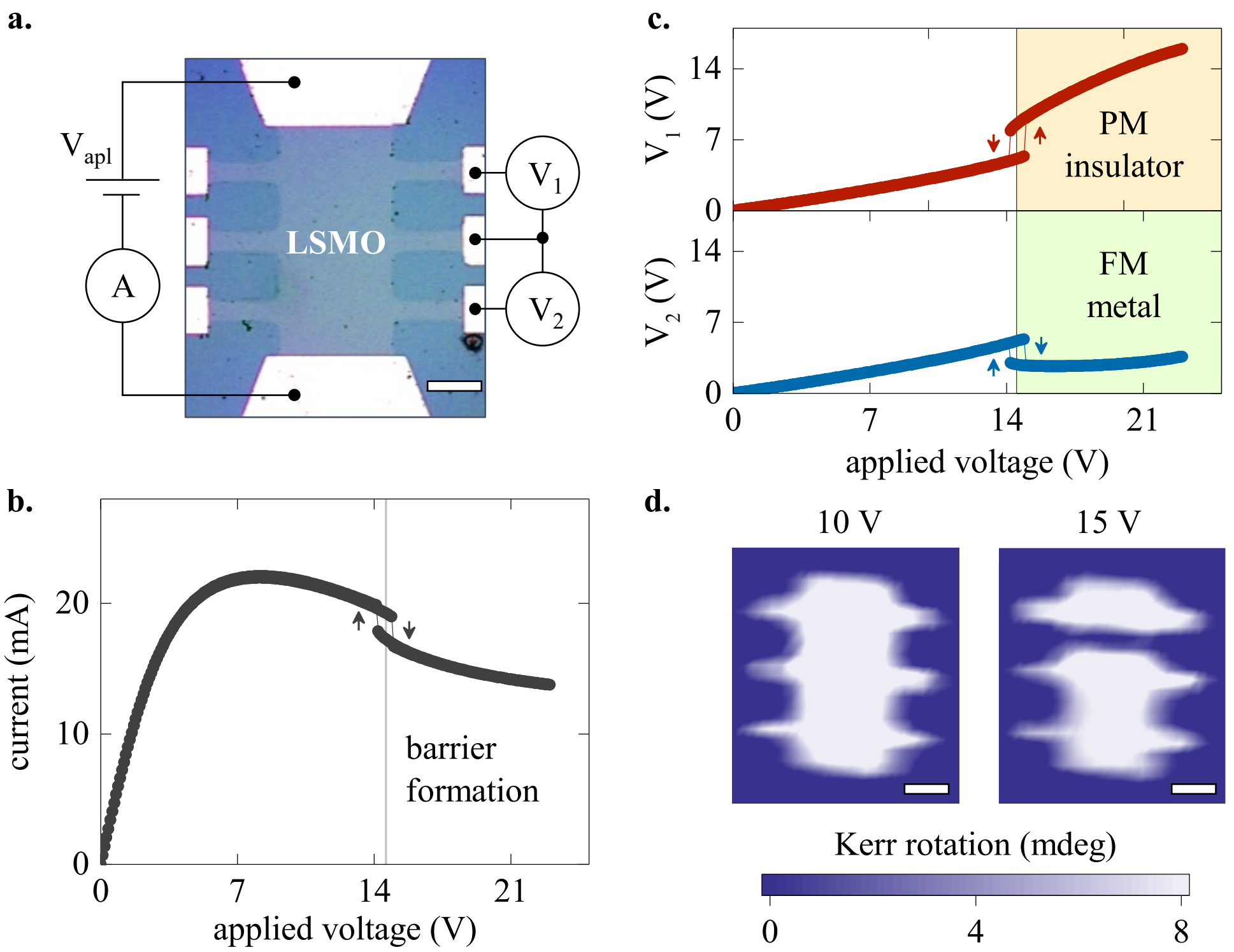

**Fig. 1. a,** Schematic of the resistive switching measurement setup. Optical image shows a 50×100 μm$^2$ LSMO Hall bar device. Scale bar is 25 μm. **b,** Global *I-V* curve of the LSMO Hall bar showing the volatile low-to-high resistance switching. **c,** Local voltages across the Hall bar arms recorded during the switching. At the 14.5 V threshold voltage, local voltage $V_1$ across the top Hall bar section abruptly increases indicating the low-to-high resistance switching. **d,** MOKE maps of the Hall bar recorded under the applied voltage of 10 V (left) and 15 V (right). While the 10 V map is uniformly ferromagnetic, the 15 V shows the formation of a PM barrier in the top section indicating the local MIT triggering. Scale bars are 25 μm. All measurements were performed at 100 K.

## II. METAL-INSULATOR TRANSITION SWITCHING

We studied the impact of electrical MIT triggering on the magnetotransport properties using the standard Hall bar geometry devices (optical image in Fig. 1a). The devices were patterned in a 20-nm-thick LSMO film epitaxially grown on a (001)-oriented $SrTiO_3$ substrate [48,49]. The Hall bars were aligned with the [100] crystallographic direction. Under equilibrium conditions, i.e., at voltages/currents far below the switching threshold, the resistance-temperature measurements showed the ferromagnetic (FM) metal to paramagnetic (PM) insulator phase transition at $T_c \sim 345$ K (Suppl. Fig. S1). In the low-temperature FM phase, the devices exhibited a $\Delta R/R \sim 0.2\%$ AMR (Suppl. Fig. S2). Near $T_c$, the application of a high magnetic field induced CMR, $\Delta R/R \sim 23\%$ at 20 kOe (Suppl. Fig. S1). Overall, the equilibrium electrical and magnetic properties of the patterned devices were typical of LSMO thin films, indicating that the device fabrication had no noticeable influence on the chemical or structural properties of the synthesized films.

The application of a large voltage triggers the MIT in LSMO devices, which produces volatile low-to-high resistance switching. Fig. 1a shows the switching measurement setup schematic. A voltage source and an ammeter (depicted on the left of the device) probed the *I-V* characteristics of the entire Hall bar. The recorded *I-V* curve (Fig. 1b) has an N-type negative differential resistance region, a part of the *I-V* curve where the slope $dV/dI < 0$, which is a signature of the low-to-high resistance switching. The switching is volatile: turning off the applied voltage automatically restores the initial low-resistance state. Metal-to-insulator switching in LSMO occurs in a characteristic spatial pattern: the formation of a localized insulating barrier perpendicular to the electric current flow [45], in contrast to the conventional conducting filament percolation along the current [50–52]. Two independent voltmeters (shown on the right of the device in Fig. 1a) monitored the local voltages across the two Hall bar sections, which allowed electrical probing of the barrier formation (Fig. 1c). Initially as the applied voltage $V_{apl}$ is ramped up, the local voltages, $V_1$ and $V_2$, are nearly equal and increase monotonically. When the MIT is triggered and the insulating barrier forms at $V_{apl} \sim 14.5$ V, the local voltage $V_1$ (top Hall bar section in Fig. 1a) abruptly increases because the barrier acts as a voltage divider focusing a large fraction of the applied voltage on itself. Consequently, the local voltage $V_2$ across the metallic Hall bar section (bottom section in Fig. 1a) decreases abruptly. By comparing $V_1$ and $V_2$, it is always possible to unambiguously identify which section of the Hall bar, top or bottom, undergoes the electrically triggered MIT and which section remains in the metallic state.

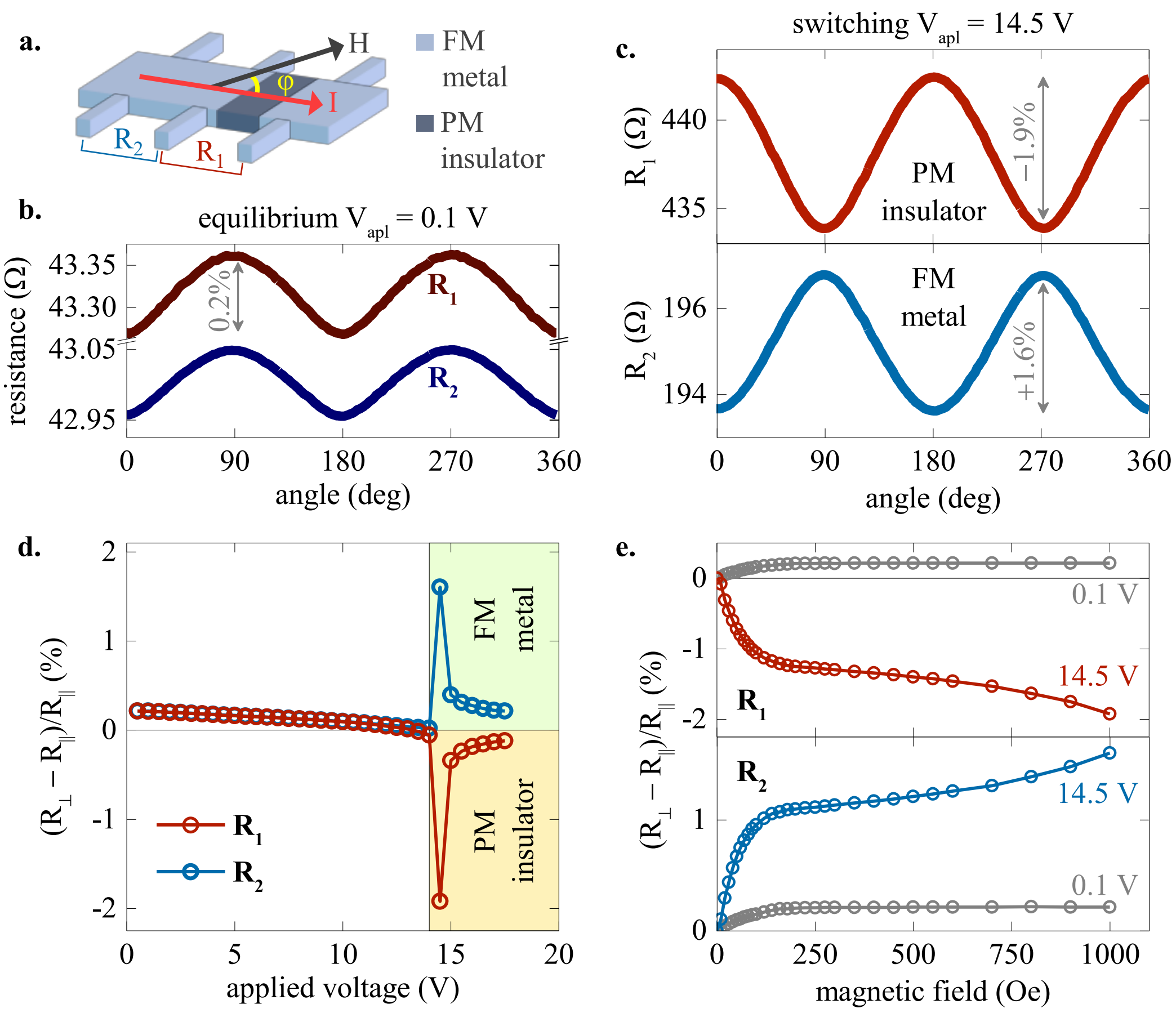

**Fig. 2.** **a**, Schematic of the AMR measurement setup. **b,** Resistance-angle curves in equilibrium (0.1 V) showing the AMR of $\Delta R/R$ ~ 0.2%. **c,** Resistance-angle curves recorded at the MIT triggering threshold (14.5 V) in the Hall bar sections containing the PM barrier (top) and in the FM section (bottom). The AMR magnitude increases and its sign flips in the PM barrier section. **d,** AMR magnitude as a function of applied voltage. The AMR anomaly can be observed at the MIT triggering threshold. AMR measurements in (**b**)-(**d**) were performed in 1 kOe applied field. **e,** AMR amplitude dependence on the applied magnetic field in equilibrium (grey lines) and at the MIT triggering threshold (red and blue lines) for the PM barrier (top) and FM metal (bottom) Hall bar sections. At the switching, the AMR amplitude shows a persistent increase with increasing field. All measurements were done at 100 K.

*In-operando* magneto-optical Kerr effect (MOKE) imaging confirmed that the abrupt increase/decrease of $V_1/V_2$ is a direct consequence of the PM insulating barrier formation. The imaging procedure (details can be found elsewhere [45]) mapped the spatial distribution of the FM regions. Because the MIT and FM transitions in LSMO are coupled [53], the MOKE maps reveal the device areas that locally undergo MIT switching (areas where the MOKE signal vanishes) and the areas that remain in the FM metal state (areas where the MOKE signal is strong). The left panel in Fig. 1d shows a MOKE map acquired at $V_{apl}$ = 10 V, i.e., a voltage that does not trigger the MIT. The entire device is uniformly FM implying that LSMO is in the metallic state. At $V_{apl}$ = 15 V (Fig. 1d, right panel), the MOKE map shows that the FM signal vanishes in the top Hall bar section. The map directly identifies the area where LSMO locally undergoes the phase transition and reveals that this phase transition results in the formation of a PM insulating barrier, i.e., a region that stretches across the full device width and blocks the current flow. The barrier appears in the same section where the local voltage $V_1$ abruptly increases in the switching *I-V* measurements (Fig. 1c). Because the transverse barrier is not expected to produce significant longitudinal current flow inhomogeneities, the total current and local voltages $V_1$ and $V_2$ can be used to calculate the effective resistances of the two Hall bar sections. Such simple local resistance calculations are justified when the switching occurs by the transverse barrier formation, while a different approach is needed when the switching causes a longitudinal filamentary percolation, which inevitably leads to a highly non-uniform current distribution [54].

## III. ANISOTROPIC MAGNETORESISTANCE

Intuitively, one might expect that the AMR will be suppressed when the applied voltage triggers the MIT because part of the device is in the PM phase and there is no AMR in the PM phase in equilibrium (Suppl. Fig. S2). Contrary to this expectation, we found that triggering the MIT leads to a strong magnetoresistance anomaly. Fig. 2b compares the AMR of the LSMO device under equilibrium conditions ($V_{apl}$ = 0.1 V) and during resistive switching ($V_{apl}$ = 14.5 V). The measurements were performed at 100 K and in 1 kOe magnetic field sufficient to saturate the magnetization in any in-plane direction. In equilibrium, the device shows conventional $\Delta R \sim sin^2\varphi$ AMR behavior with the magnitude of $(R_{\perp} - R_{||})/R_{||} \sim +0.2\%$, where $R_{\perp}$ and $R_{||}$ are the resistances when the magnetization is saturated in the in-plane direction perpendicular or parallel to the electric current, respectively (Fig. 2a). When applied voltage triggers the MIT at 14.5 V (Fig. 2c), the AMR magnitude in the Hall bar section that remains in the FM state increases by a factor of ~8 reaching $(R_{\perp} - R_{||})/R_{||} \sim +1.6\%$. In the Hall bar section that has the PM barrier, the AMR magnitude also increases and, surprisingly, flips sign becoming $(R_{\perp} - R_{||})/R_{||} \sim -1.9\%$. The two $R(\varphi)$ curves (FM metal and PM barrier sections) appear nearly as mirror reflections of each other. This reciprocal symmetry of the $R(\varphi)$ curves is an indication that the coupling between the two Hall bar sections during the MIT switching, i.e., the resistance change of one section is reflected in the other section, is responsible for the anomalous magnetotransport properties, as discussed later in the paper.

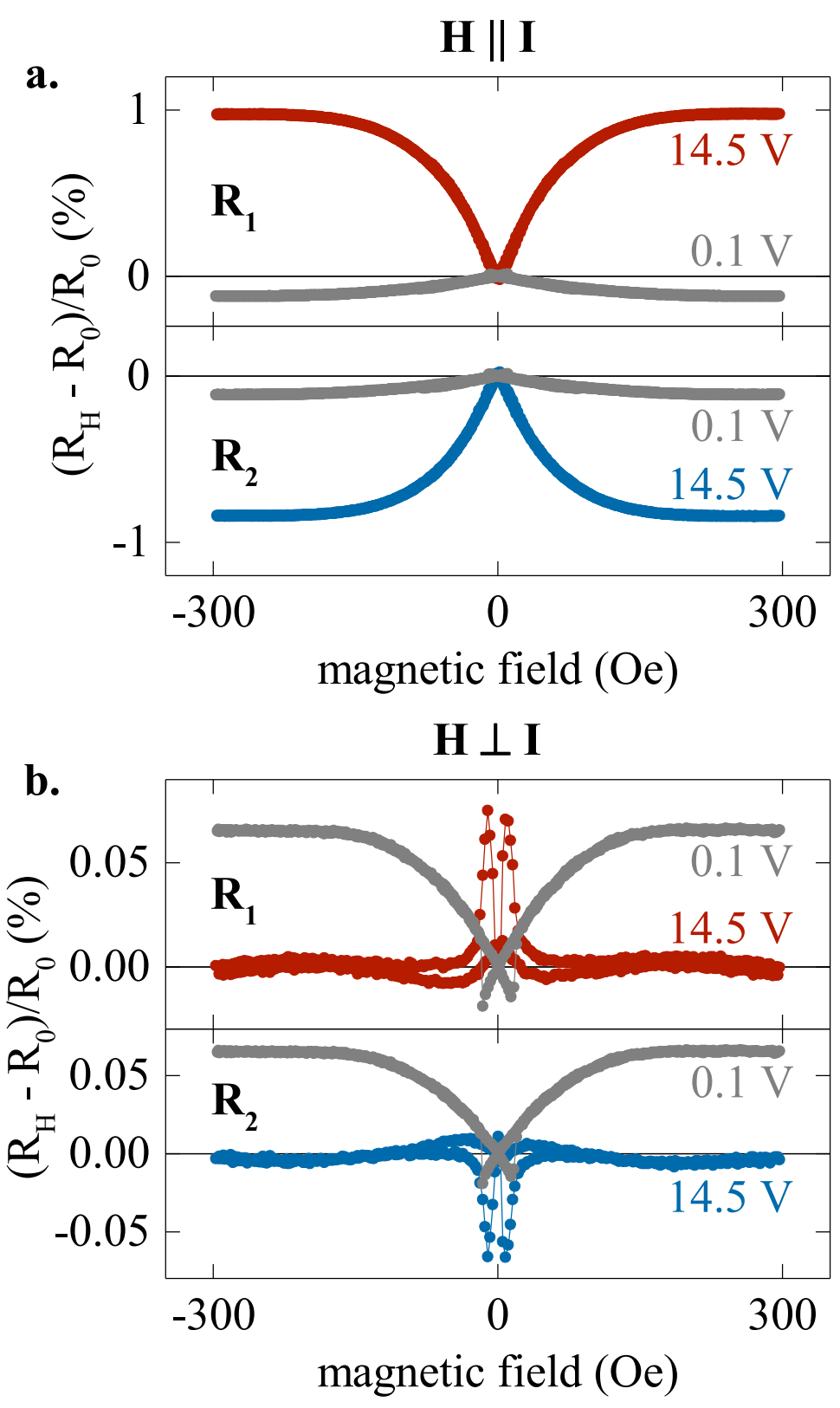

**Fig. 3.** Small-field magnetoresistance measurements with the magnetic field parallel (**a**) and perpendicular (**b**) to the electric current in equilibrium (0.1 V, gray lines) and at the MIT triggering threshold voltage (14.5 V, red and blue lines) corresponding to the two sections of the Hall bar device, $R_1$ and $R_2$ (defined in Fig. 2a). During resistive switching, the magnetoresistance of the PM barrier section ($R_1$) flips sign. The shape and magnitude differences between $R_{||}$-$H$ and $R_{\perp}$-$H$ curves indicate the change of magnetic anisotropy during the switching. All measurements were performed at 100 K.

The AMR anomaly develops abruptly at the MIT triggering threshold voltage. Fig. 2d shows the voltage dependence of AMR magnitude of the two Hall bar sections, one that undergoes switching into the PM insulator state (red curve) and the other one that remains in the FM metal state (blue curve). Before the switching is induced at 14.5 V, the AMR in both sections is equal and decreases slightly with increasing voltage, which most likely originates from uniform Joule heating that precedes the switching [45]. A sharp peak and the sign flip of $(R_{\perp} - R_{||})/R_{||}$ in the barrier section coincides with the threshold voltage that triggers the switching and causes the PM insulating barrier formation. At the threshold voltage, the barrier is the narrowest [45]. As the voltage increases, the barrier width grows and, simultaneously, the magnitude of the AMR anomaly decreases. The sign flip of $(R_{\perp} - R_{||})/R_{||}$, however, persists in the PM barrier section as long as the device is in the switched state.

Electrical MIT triggering in LSMO is mediated by Joule heating [45]. Even though the switching process is thermal in nature, the AMR magnitude voltage dependence drastically differs from the equilibrium temperature dependence (compare Fig. 2d and Suppl. Fig. S2): (1) equilibrium AMR magnitude never exceeds 0.25%, while the AMR reaches ~1.9% during switching, (2) $R_{\perp} > R_{||}$ at any temperature in equilibrium, while $R_{\perp} < R_{||}$ in the PM barrier section during switching. These two facts highlight the importance of the voltage induced PM barrier formation in the emergence of magnetoresistance anomaly since simple Joule heating cannot account for the experimental observations.

AMR is expected to be independent of the applied magnetic field magnitude provided that the field is sufficient to saturate the magnetization along all in-plane directions. This expectation is indeed satisfied in the LSMO devices in equilibrium, e.g., $(R_{\perp} - R_{||})/R_{||}$ has the same value of ~0.2% at 200 Oe and at 1000 Oe fields (Fig. 2e, gray lines). During the switching, the AMR magnitude increases monotonically with increasing magnetic field without any signs of saturation (Fig. 2e, color lines). Because the AMR anomaly is largest near the MIT triggering voltage (Fig. 2d), the absence of AMR saturation is likely an indication that the applied magnetic field drives the LSMO toward the state close to the switching threshold, i.e., when the barrier is narrowest, which could be due to the CMR mediated voltage/power redistribution, as discussed later in the paper.

Small-field magnetoresistance measurements showed evidence of the magnetic anisotropy change when the LSMO device undergoes the MIT switching. Fig. 3 compares the resistance-field curves acquired with the magnetic field parallel, $R_{||}$-$H$, and perpendicular, $R_{\perp}$-$H$, to the electric current. A linear CMR contribution was subtracted from the $R$-$H$ curves in order to highlight the anisotropy related effects. Raw magnetoresistance data are shown in Suppl. Fig. S3. In equilibrium, the $R_{||}$-$H$ and $R_{\perp}$-$H$ curves (Fig. 3, grey lines) have opposite signs due to the intrinsic AMR. Both $R$-$H$ curves, however, have similar shapes and comparable magnitudes (~0.1%), which is indicative of the conventional easy-plane magnetic anisotropy in a thin film. When the applied voltage triggers the MIT, the $R$-$H$ curves of the FM metal and PM barrier Hall bar sections (Fig. 3, red and blue lines) have the opposite signs consistent with the AMR sign flip discussed in the previous paragraphs (Fig. 2 c-e). Importantly, the $R_{||}$-$H$ and $R_{\perp}$-$H$ curves have drastically different shapes and magnitudes during the switching. $R_{||}$-$H$ curve displays a gradual evolution without distinct magnetization reversal features and it has the large magnitude of ~1%. Conversely, the magnetization reversal events are strongly pronounced in the $R_{\perp}$-$H$ curves, while $R_{\perp}$ has nearly the same value at remanence (zero field) and at the maximum applied field (300 Oe). The observed differences between the $R_{||}$-$H$ and $R_{\perp}$-$H$ curves suggest that during the switching, the magnetization prefers the direction perpendicular to the current (i.e., aligned with the transverse PM barrier), consistent with our recent MOKE studies of the uniaxial anisotropy development in the LSMO switching devices [47].

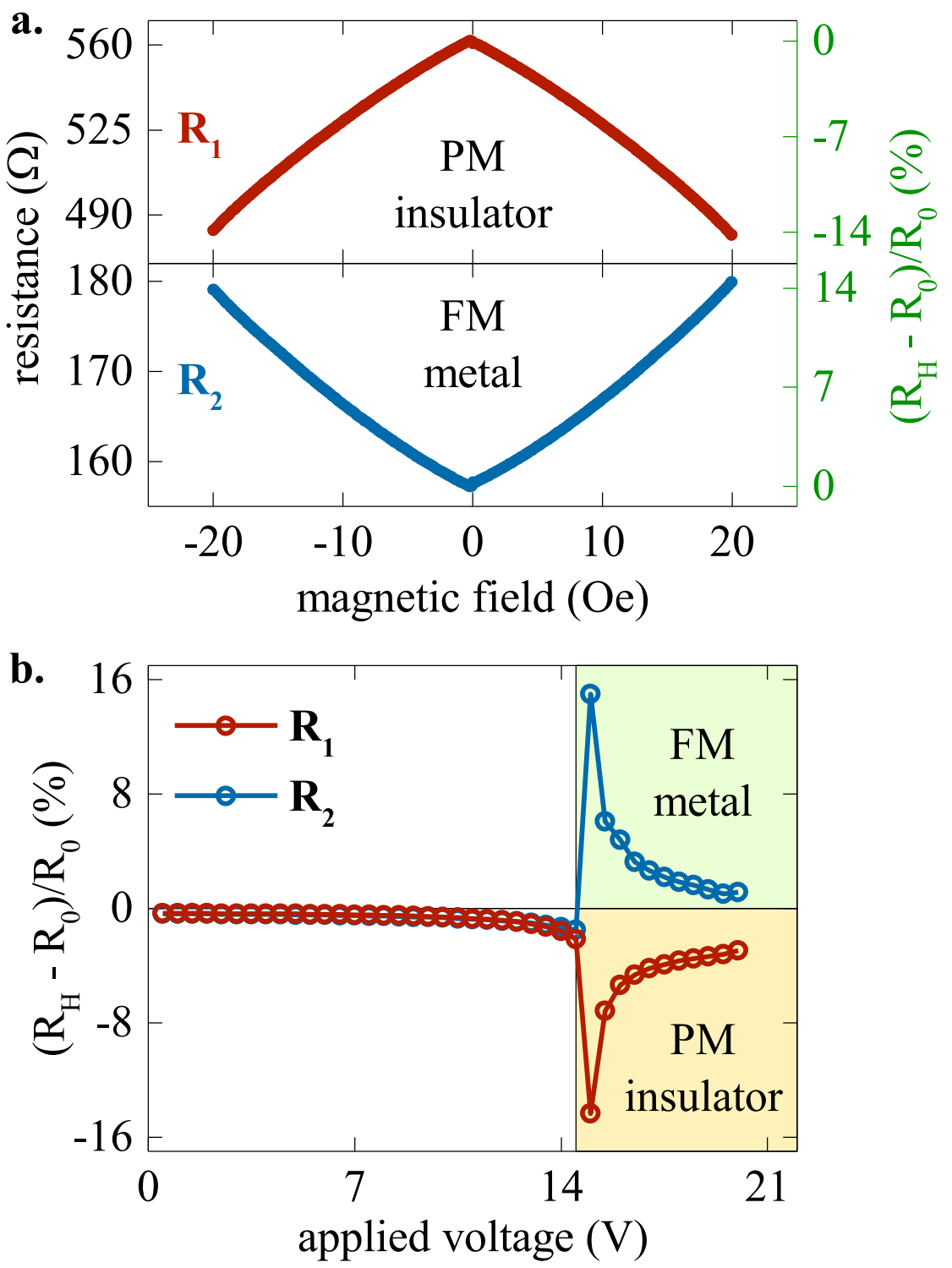

**Fig. 4. a,** Large-field magnetoresistance measurements at the MIT triggering threshold voltage (14.5 V). The *R-H* curve of the PM barrier section of the Hall bar (top, red line) exhibits conventional negative CMR. The *R-H* curve of the FM metal section shows an anomalous positive CMR. **b,** Magnetoresistance at 20 kOe field as a function of the applied voltage. Similar to AMR in Fig. 2, CMR shows a strong anomaly at the MIT threshold triggering voltage. $R_1$ and $R_2$ notations are defined in Fig. 2a. All measurements were performed at 100 K.

## IV. COLOSSAL FMAGNETORESISTANCE

An anomaly coinciding with the electrical triggering of MIT also appears in CMR. In equilibrium, the application of a high magnetic field reinforces the FM metal phase in LSMO, which leads to a large (i.e., colossal) resistance decrease (negative magnetoresistance) near the FM metal to PM insulator phase transition [53]. Because the electrical MIT triggering in LSMO is mediated by Joule heating [45], it may be expected that the regions near the PM barrier will locally exhibit a CMR as their temperature is close to $T_c$, while the magnetoresistance will be small in the FM regions as their temperature is far below $T_c$. Contrary to the expectations, we observed a strong magnetoresistance response both in the PM barrier and FM metal Hall bar sections during the switching (Fig. 4). Magnetoresistance in the PM barrier section is negative reaching –15% at 20 kOe field (Fig. 4a, top panel). This PM barrier magnetoresistance is comparable to the equilibrium CMR at $T_c$, -15% vs. -23%, respectively, although only a small volume fraction of the device is expected to be close to $T_c$ at the switching threshold voltage, while the entire material volume contributes to the equilibrium CMR.

Surprisingly, the FM metal section also exhibits an anomalous CMR-like effect, but its magnetoresistance is positive, +15% (Fig. 4a, bottom panel), which has never been reported in LSMO under equilibrium conditions. We note that although the relative magnetoresistance values of the PM barrier and FM sections have nearly equal magnitudes, –15% vs. +15%, the absolute resistances differ by the factor of ~3, allowing the unambiguous identification of the Hall bar section that undergoes the MIT switching. The two *R-H* curves (PM barrier and FM metal) appear nearly as mirror reflections of each other, which suggests that the MIT switching is responsible for the anomalous CMR properties, similar to the previously discussed AMR anomaly.

The strongest CMR anomaly occurs at the MIT triggering threshold (Fig. 4b). At the threshold voltage, the I-V curve has the sharpest slope (Fig. 1b) corresponding to a rapid resistance change, i.e., the resistive switching. Consequently, even a small external stimulus that influences the power/voltage distribution balance in the device can drive a large resistive change response. As we discuss next, the resistance modulation associated with the intrinsic AMR and CMR can couple with the resistive switching process, which leads to an emergent magnetoresistance anomaly.

## V. SWITCHING MODEL

To investigate the possible mechanism behind the observed AMR and CMR anomalies, we developed a hybrid model by combining a resistive network with the Ising model (Fig. 5a). The value of each resistor in the network depends on magnetization $m$ and temperature $T$:

$$R(m,T) = R_0 T e^{\varepsilon \frac{1-m^2}{T}} \tag{1}$$

where $R_0$ and $\varepsilon$ are constants. Eq. (1) was found to provide a good phenomenological description of the MIT and CMR in rare-earth manganites [55,56]. In the simulations, we set the applied voltage and magnetic field and we solve the resistor network by considering the electro-thermal effects [45,54]. Using local temperature, we compute local magnetization from the mean-field Ising model solution:

$$m = \tanh \frac{h - T_c m}{T} \tag{2}$$

where $h$ is the magnetic field and $T_c$ is the critical temperature. The obtained local magnetization and temperature are then used to update the resistor values in the network. For each applied voltage and field, we ran an iterative process until all variables converge.

Similar to the experimental results (Fig. 1d), the model shows that the switching occurs by the formation of a high-temperature insulating barrier inside the low-temperature metallic matrix (Fig. 5 b and c). The model further yields that the barrier's resistance and size depend on the applied magnetic field (Fig. 5d). As the field increases, the barrier region (whose temperature is just above $T_c$) develops a noticeable magnetization (Fig. 5c) leading to a resistance decrease given by Eq. (1), i.e., the intrinsic CMR. This resistance decrease in the high-temperature barrier region lowers the total resistance of the entire device causing the power dissipation and temperature increase because the device is under a constant applied voltage (Fig. 5e). As the barrier temperature is close to $T_c$, the intrinsic CMR produces a larger resistance change compared to the effect of the temperature change induced by the applied magnetic field, which explains the negative magnetoresistance of the barrier Hall bar section observed in the experiments (Fig. 4). The temperature of metallic matrix, on the other hand, is much lower than $T_c$. The metallic matrix, consequently, does not exhibit a noticeable intrinsic CMR leaving the magnetic field induced temperature increase as the primary factor determining the resistance of the metal regions. Due to the positive resistance-temperature dependence in the metallic state, the resistance of the metallic matrix increases with magnetic field (inset in Fig. 5b), which explains the positive magnetoresistance of the metallic Hall bar section observed in the experiments (Fig. 4).

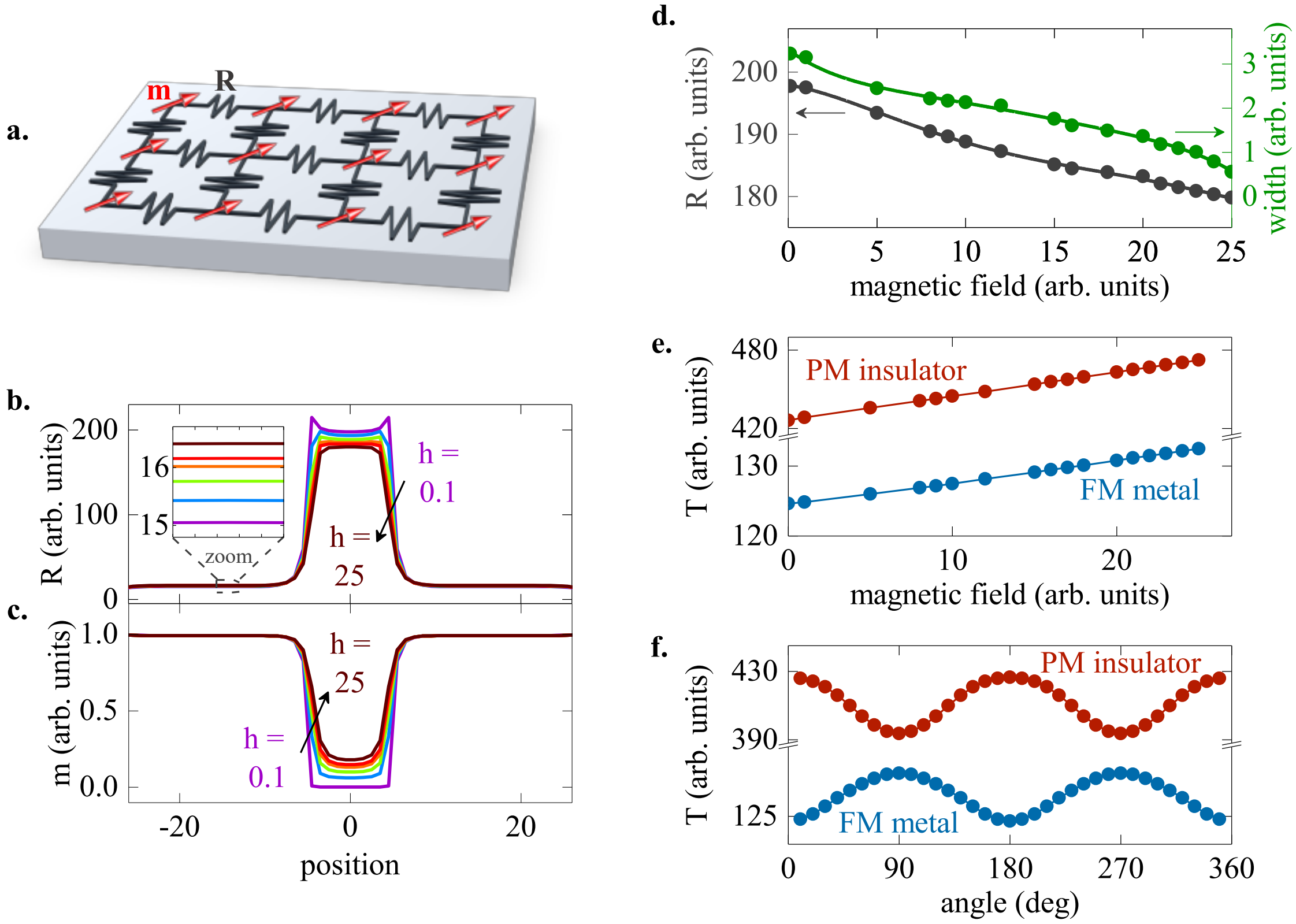

**Fig. 5. a,** Schematic of the hybrid resistor network + Ising model. **b, c,** Magnetic field dependence of the resistance (**b**) and magnetization profiles (**c**) across the switching device length at an applied voltage above the MIT triggering threshold. The formation of a PM insulating barrier at the device center is clearly visible. The inset in (**b**) shows a zoom-in of the resistance profile in a metallic region of the device. The resistance of the metallic regions increases with increasing magnetic field, which yields a positive magnetoresistance. **d,** Magnetic field dependence of the resistance (black curve) and width (green curve) of the PM barrier. **e,** Temperature of the PM barrier (red curve) and metallic matrix (blue curve) as a function of the magnetic field. **f,** Temperature of the PM barrier (red curve) and metallic matrix (blue curve) as a function of the angle between the magnetization and current.

To explore the effect of AMR during the switching, we explicitly introduced the resistance angular dependence on the magnetization direction into Eq. (1) by setting

$$R_0(\varphi) = R_{||} + (R_{\perp} - R_{||}) \sin^2 \varphi \tag{3}$$

We fit the experimental AMR data (Suppl. Fig. 2) to obtain the temperature-dependent $R_{||}$ and $R_{\perp}$ values used in the simulations. The model revealed that in the switched state, temperature of both the metallic matrix and insulating barrier depends on the magnetization direction (Fig. 5f). This angular dependence of temperature can be understood as following. The metallic matrix has the intrinsic positive AMR, i.e., its resistance is maximized when the magnetization is perpendicular to the current ($R_{\perp} > R_{||}$). In the resistive switching state, a small AMR-induced resistance increase leads to an increase of the dissipated power (hence, the temperature increase) in the metallic region because of the voltage focusing. As the AMR initiates the voltage/power focusing, this focusing is further amplified by the natural positive resistance-temperature dependence in the metallic state. Thus, the combined effect of the intrinsic AMR and resistance-temperature dependence can explain the effective AMR increase in the metallic section of the Hall bar observed in the experiments (Figs. 2 and 3). On the other hand, as the metallic regions focus more dissipated power, the temperature of the barrier section decreases (Fig. 5f) and, consequently, its resistance also decreases due to the natural resistance-temperature dependence. This decrease in resistance produces an apparent negative AMR observed in the barrier section (Fig. 2) even though the barrier itself is in the PM state and does not have an intrinsic AMR.

Overall, the model shows that the coupling between the intrinsic magnetoresistance and resistive switching state leads to an anomalous CMR and AMR. During the electrical triggering of MIT, a phase separation pattern of the insulating and conducting regions emerges. The size and temperature of these regions are highly susceptible to external stimuli influence because the phase separation pattern depends on a delicate voltage/power distribution balance in the device. Even a small resistance change produced by the intrinsic magnetoresistance can significantly alter this balance resulting in the temperature and size change of the insulating and conducting regions. An effective magnetoresistance amplification can be achieved when the intrinsic magnetoresistance and voltage/power redistribution act together, i.e., both effects cause a resistance change in the same direction.

## VI. CONCLUSION

This work reports an anomalous magnetoresistance behavior in LSMO devices when the material is driven into the resistive switching state. At the MIT triggering threshold voltage, both the AMR and CMR show a large increase in magnitude. Surprisingly, an effective AMR increase and a sign flip can be observed in the device regions switched into the PM insulating phase even though the PM phase does not have an AMR in equilibrium. The FM metal matrix surrounding the switched regions has a strong positive CMR response, in stark contrast to the intrinsic negative CMR that appears near $T_c$ under equilibrium conditions. We attribute the above effects to the coupling between the intrinsic magnetoresistance and voltage/power distribution in the resistive switching state. This coupling enables the size and temperature control of the metal/insulator regions by applying an external magnetic field. The key feature producing the magnetoresistance anomaly is the emergence of a characteristic metal/insulator phase separation pattern during the electrical MIT triggering, specifically for LSMO, the formation of an insulating barrier inside the metallic matrix. Percolating filaments or blocking barriers are commonly observed in resistive switching devices [45,51,52]. Therefore, it can be expected that unusual magnetotransport phenomena might be also found in a variety of magnetically ordered MIT switching systems, such as other magnetic members of the rare-earth manganite family [53,57], antiferromagnetic rare-earth nickelates [58,59], ferrimagnetic $Fe_3O_4$ [60,61], and antiferromagnetic $V_2O_3$ [51,62].

### Acknowledgments

The magnetism research was supported by the U.S. Department of Energy (DOE), Office of Science, Basic Energy Sciences (BES), Materials Sciences and Engineering Division under Award # DE-FG02-87ER45332. Material synthesis, characterization and numerical simulation were supported as a part of Quantum Materials for Energy Efficient Neuromorphic Computing (Q-MEEN-C), an Energy Frontier Research Center funded by the DOE, Office of Science, BES under Award # DE-SC0019273

# Magnetoresistance anomaly during the electrical triggering of a metal-insulator transition

Pavel Salev[1], Lorenzo Fratino[2,3], Dayne Sasaki[4], Soumen Bag[2], Yayoi Takamura[4],
Marcelo Rozenberg[2], Ivan K. Schuller[5]

*[1]Department of Physics & Astronomy, University of Denver, Denver, CO*
*[2]Université Paris-Saclay, CNRS Laboratoire de Physique des Solides, 91405, Orsay, France*
*[3]Laboratoire de Physique Théorique et Modélisation, CNRS UMR 8089, CY Cergy Paris Université, 95302 Cergy-Pontoise Cedex, France*
*[4]Department of Materials Science and Engineering, University of California Davis, Davis, CA*
*[5]Department of Physics and Center for Advanced Nanoscience, University of California San Diego, La Jolla, CA*

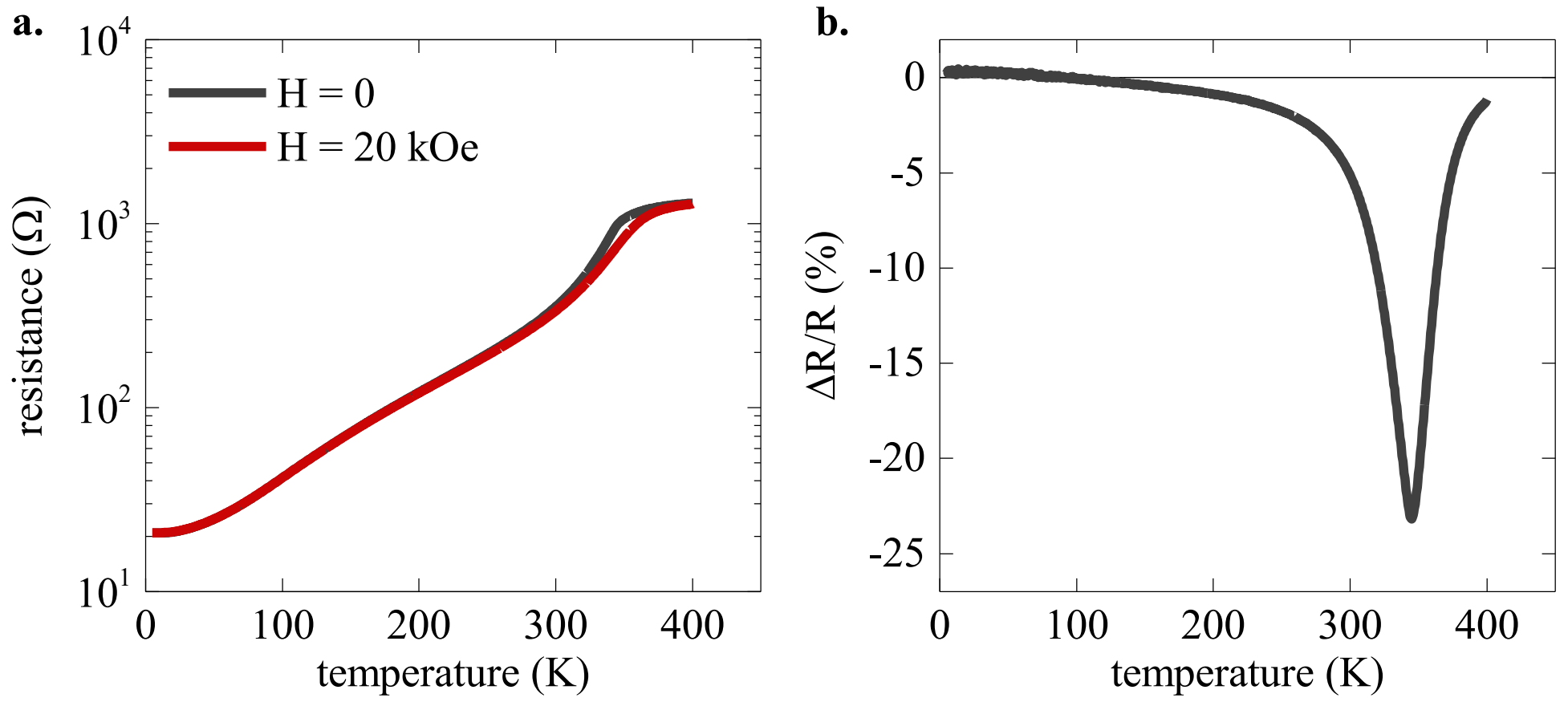

**Suppl. Fig. S1. a,** Resistance-temperature dependence of the patterned LSMO Hall bar sample in zero field (grey line) and in 20 kOe field (red line). **b**, Resistance ratio between the 20 kOe and zero field curves from the panel **a** showing the colossal magnetoresistance (CMR) of about -23% at $T_c$ = 345 K. CMR is defined with respect to the resistance in zero magnetic field.

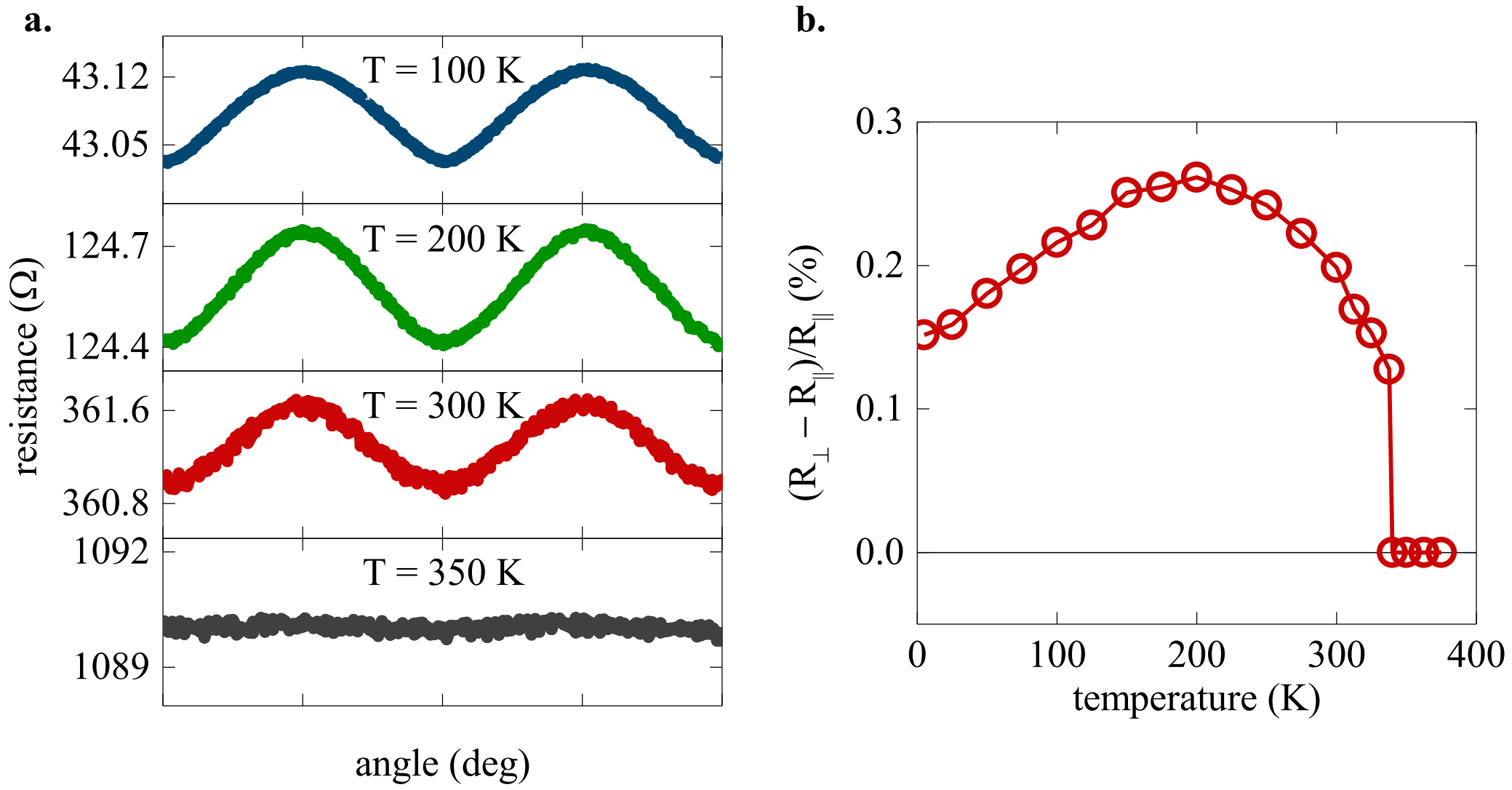

**Suppl. Fig. S2. a,** Examples of the equilibrium anisotropic magnetoresistance (AMR) measurements at a few select temperatures. Above $T_c$ = 345 K, the AMR vanishes. The measurements were done in 1 kOe field. **b**, AMR magnitude (difference between the current perpendicular and current parallel to magnetization) as a function of temperature. AMR magnitude is defined with respect to the resistance when magnetization is saturated along the direction of the electric current flow.

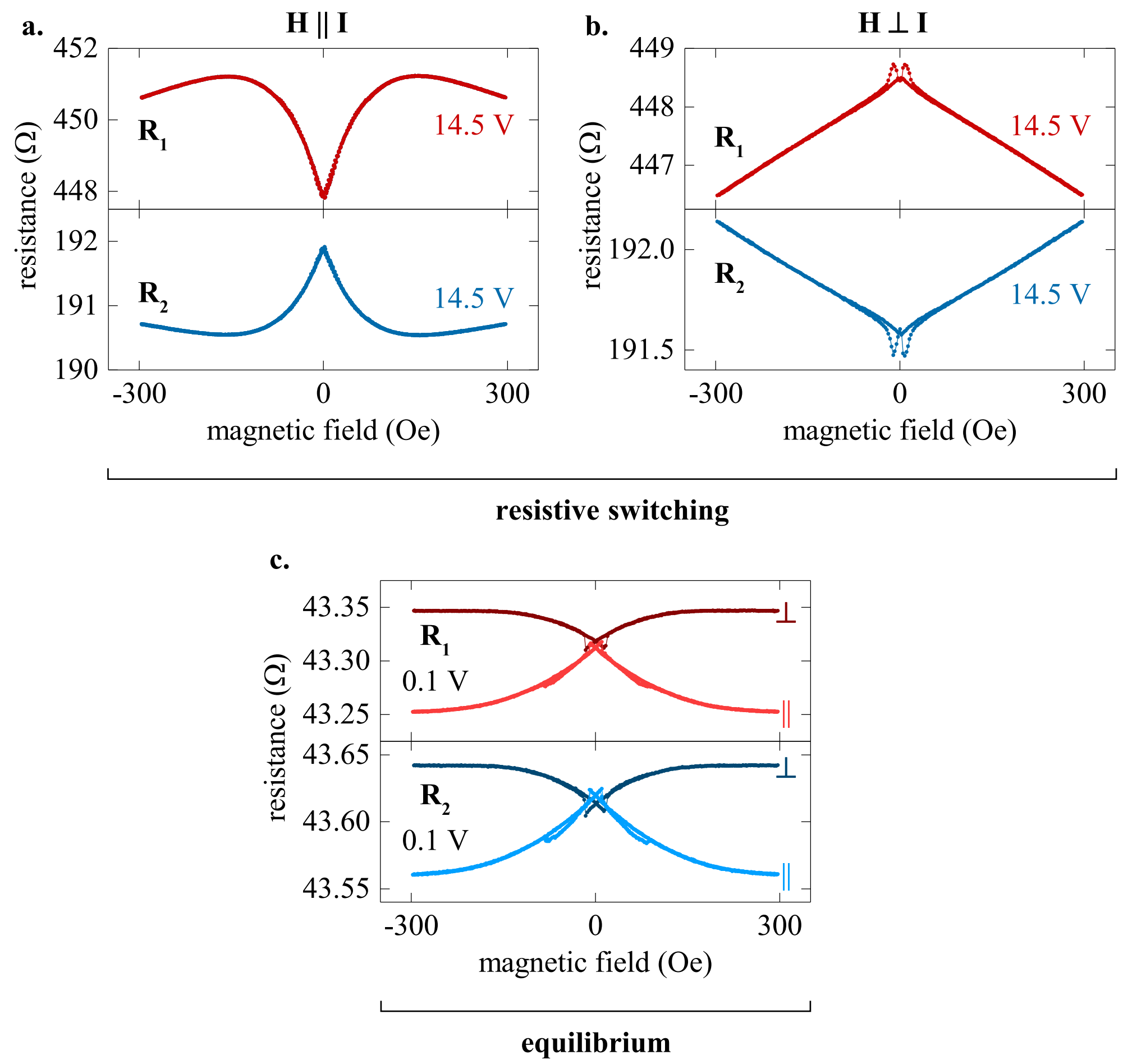

**Suppl. Fig. S3.** Raw magnetoresistance data that was used to plot Fig. 3 in the main text. Resistances $R_1$ and $R_2$ correspond to the two sections of the Hall bar (defined in the main text Fig. 2). **a**, Resistance-field curves in the current **parallel** to magnetization configuration under the application of 14.5 V (MIT triggering threshold). The top curve ($R_1$) corresponds to the section of the Hall bar that contains the paramagnetic insulating barrier (high resistance). The bottom curve ($R_2$) corresponds to the ferromagnetic metal Hall bar section (low resistance). **b,** Resistance-field curves in the current **perpendicular** to magnetization configuration under the application of 14.5 V (MIT triggering threshold). The top curve ($R_1$) corresponds to the section of the Hall bar that contains the paramagnetic insulating barrier (high resistance). The bottom curve ($R_2$) corresponds to the ferromagnetic metal Hall bar section (low resistance). **c,** Equilibrium resistance-field curves of the two Hall bar sections ($R_1$ and $R_2$) in the current **parallel** (light color lines) and current **perpendicular** (dark color lines) to magnetization configurations. All measurements were done at 100 K.